\documentclass[prb,aps,twocolumn,floatfix]{revtex4-1}
\usepackage{textcomp}
\usepackage{graphicx}
\usepackage{amssymb}
\usepackage{epsfig}
\usepackage{gensymb}

\begin{document}

\title{Molybdenum-Rhenium superconducting suspended nanostructures}

\author{Mohsin Aziz}
\author{David Christopher Hudson}
\author{Saverio Russo}
\affiliation{Centre for Graphene Science, College of Engineering, Mathematics and Physical Sciences, University of Exeter, Exeter EX4 4QF, UK}

\begin{abstract}
Suspended superconducting nanostructures of MoRe $50\%/50\%$ by weight are fabricated employing commonly used fabrication steps in micro- and nano-meter scale devices followed by wet-etching with Hydro-fluoric acid of a SiO$_2$ sacrificial layer. Suspended superconducting channels as narrow as $50\,\rm{nm}$ and length $3\,\rm{\mu m}$ have a critical temperature of $\approx 6.5\,\rm{K}$, which can increase by $0.5\rm{K}$ upon annealing at $400\,^{\circ}\mathrm{C}$. A detailed study of the dependence of the superconducting critical current and critical temperature upon annealing and in devices with different channel width reveals that desorption of contaminants is responsible for the improved superconducting properties. These findings pave the way for the development of superconducting electromechanical devices using standard fabrication techniques.
\end{abstract}

\maketitle

Advances in fundamental science are sometimes only possible with adequate technological progress, allowing access to unprecedented experimental conditions necessary to verify the validity of theoretical assumptions. For example, the ability to controllably fabricate high quality multilayer structures of hybrid superconductor/normal metal materials with ultra-thin superconductors led to the theoretically unexpected \cite{Falci2001} observation of crossed-Andreev reflection \cite{Russo2005}. Similarly, the coupling between the electron degree of freedom and the mechanical vibration of nanometer-scale systems has been the object of extensive studies \cite{Book1,Science1,Science2,Science3}. The interplay between correlated electrons in a superconductor and mechanical vibrations grants access to unique physical scenarios, such as cooling the nanomechanical motion of a resonator to its groundstate \cite{Lehnert2008}.  

Nano- and micro-electromechanical structures are typically fabricated on a sacrificial substrate, which is subsequently removed in a final etching step leaving a suspended device free to vibrate \cite{Fabrication1}. The choice of the material to be used as a sacrificial layer is dictated by the compatibility between acids (solvents or gases in reactive ion etching) needed to etch this layer and the ability of the metallic structures to withstand the process. For example, commonly used inorganic semiconductors and oxides are only etched by harsh acids which would also etch most superconducting materials. At the same time, polymer sacrificial layers are easy to remove with mild solvents \cite{Tombros1,Du2014} but they contaminate the superconducting metals with a negative effect on the superconducting properties \cite{Ref5,Ref6,Ref7}. Reactive ion etching \cite{Lehnert2008} is another viable way to remove a sacrificial layer, however this is difficult to use with atomically thin materials such as graphene which easily etched by low power plasmas \cite{Monica}. Finding a superconducting metal compatible with standard wet-etching of commonly used oxide sacrificial layers (e.g. SiO$_2$) and for which it is easy to fabricate high quality nanostructures would broaden the horizons of the science of superconducting electromechanical devices to include emerging atomically thin systems. To date, no such a material or technology has been reported yet experimentally.

\begin{figure}[ht]{}
\includegraphics[width=0.47\textwidth]{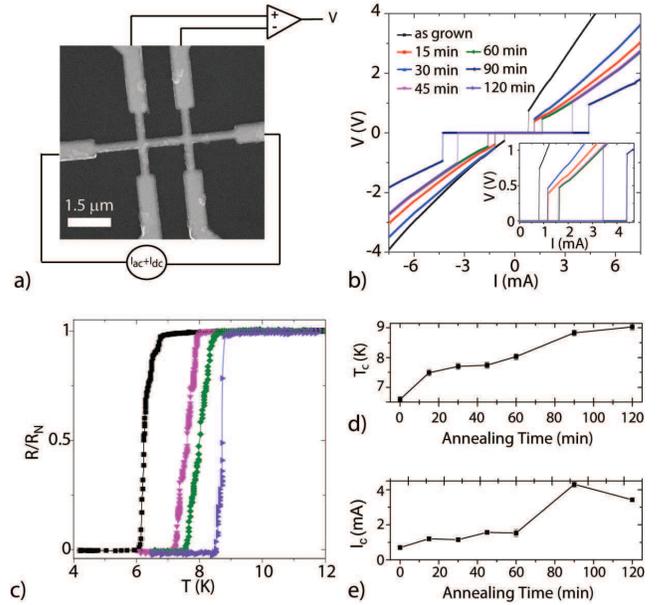}
\caption{\label{fig1} (a) Shows a scanning electron micrograph (SEM) of a Hall bar geometry ($\rm{W}\,=\,200\,\rm{nm}$) with a schematic representation of the measurement setup used. (b) is a plot of representative voltage \textit{versus} current characteristics measured in devices with $\rm{W}\,=\,200\,\rm{nm}$ after performing annealing for different durations as indicated in the legend. The inset shows a zoom in the low voltage range to highlight the values of critical current. (c) shows representative plots of the normalized 4-terminal device resistance for $\rm{W}\,=\,200\,\rm{nm}$ and for different annealing time as indicated by the legend in (b). (d-e) are summary plots of critical temperature and critical current for the measured MoRe Hall bars structures with channel width of $200\,\rm{nm}$ as a function of annealing duration, respectively.}
\end{figure}

Here we demonstrate that an alloy of MoRe $50\%/50\%$ by weight is ideally suited for the fabrication of suspended nanometer scale devices using standard lithographic techniques and Hydro-fluoric etching of SiO$_2$ sacrificial layer. Narrow channels of $200\,\rm{nm}$ width have a critical temperature as high as $\approx 9\,\rm{K}$. The superconducting properties, such as critical current and critical temperature are found to improve upon annealing the devices at $400\,^{\circ}\mathrm{C}$. A detailed study of the evolution of the superconducting properties upon annealing in devices with different channel width reveals that this annealing step desorbs polymethylmethacrylate (PMMA) contaminants from MoRe. Finally, we demonstrate that suspended superconducting nanobridges with channel width as narrow as $50\,\rm{nm}$ and length up to $3\,\rm{\mu m}$ with a critical temperature comparable to the supported devices ($\approx 7\,\rm{K}$ for channel width of $200\,\rm{nm}$ and $\approx 6.5\,\rm{K}$ for $50\,\rm{nm}$) can be fabricated using standard lithography and lift-off techniques with PMMA. These findings pave the way for the development of superconducting electromechanical devices.

The fabrication of superconducting nanostructures typically requires hard masks (e.g. Ge \cite{Ref5}) or hard baked polymers to reduce the contamination of the superconductor \cite{Ref6,Ref7}. To demonstrate the potential technological advantage of MoRe over other materials we purposefully choose one of the most common fabrication procedures in micro- and nanoelectronics, that is also one of the worst choices for fabricating superconducting nano-structures. The procedure uses standard PMMA, low baking temperature ($180\,^{\circ}\mathrm{C}$ on a hotplate) and very short baking time (just $2\,\rm{min.}$) likely to lead to high levels of contaminations of the metal by the PMMA. The devices are fabricated on substrates of Si coated by $300\,\rm{nm}$ thick SiO$_{2}$. Multi-terminal superconducting nanostructures are defined using standard electron beam lithography on a single layer of PMMA 950K A6 ($\approx 300\,\rm{nm}$ thick) and development at room temperature in 4-methylpentan-2-one:Isopropyl Alcohol=1:3 for $30\,\rm{sec}$. The deposition of the superconducting alloy MoRe is carried out by Ar sputtering in a high vacuum system with background pressure $<2 \times10^{-8}\,\rm{Torr}$, without performing any plasma or UV-cleaning of the substrate prior to the deposition. MoRe is sputtered at a rate of $28\,\rm{nm/min}$ for $4\,\rm{min.}$ and a final thickness of $115\,\rm{nm}$ is achieved. In a second lithography step normal metal contacts and bonding pads are defined using Cr/Au ($7/70\,\rm{nm}$). With this procedure, we define superconducting nano-channels of length up to $3\,\rm{\mu m}$ and widths (W) $50\,\rm{nm}$ and $200\,\rm{nm}$.   

After the devices have been fully fabricated, we perform a post-processing annealing treatment in a furnace under constant flow of Ar/H$_2$ gasses, at  $400\,^{\circ}\mathrm{C}$ and for different durations (15, 30, 45, 60, 90, and $120\,\rm{min.}$). We then characterize the superconducting properties such as critical temperature and critical current in He$_4$ atmosphere over a wide temperature range (from room temperature down to $4.2\,\rm{K}$) using constant current in a 4-terminal configuration as shown in the schematic of Figure 1a. 

Figure 1b shows the measured voltage \textit{versus} current characteristics for superconducting channels of $\rm{W}\,=\,200\,\rm{nm}$ after different steps of annealing. It is apparent that, in spite of the deliberate choice of fabrication procedure subject to a high level of contamination, in all cases we measure large values of superconducting critical current (I$_{\rm{C}}$), which exceed $700\,\rm{\mu A}$ at $4.2\,\rm{K}$. Furthermore, the specific value of I$_{\rm{C}}$ is found to increase upon increasing the annealing time, reaching values of $4.1\,\rm{mA}$ after $90\,\rm{min.}$ of annealing, and it reduces to $3.5\,\rm{mA}$ after $120\,\rm{min.}$ annealing time. At the same time, the critical temperature (T$_{\rm{C}}$) is found to increase monotonously from $\approx\,6\,\rm{K}$ in as-grown devices up to $9\,\rm{K}$ after $120\,\rm{min.}$ of annealing, see Figure 1c-d. The observed increase of T$_{\rm{C}}$ is also accompanied by a sharper normal metal-to-superconductor transition as seen in a plot of the temperature dependence of the normalized resistance to the normal metal state value (R$_{\rm{N}}$), see Figure 1c. Finally, the plots in Figure 1d-e summarize the overall evolution of I$_{\rm{C}}$ and T$_{\rm{C}}$ as a function of the annealing time up to $120\,\rm{min.}$. 

 \begin{figure}[ht]{}
\includegraphics[width=0.47\textwidth]{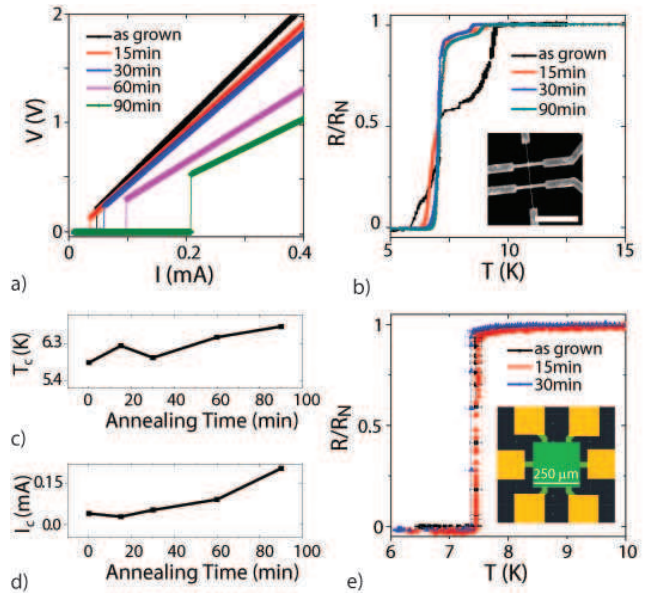}
\caption{\label{fig2} (a) is a plot of representative V-I characteristics for MoRe Hall bar of $\rm{W}\,=\,50\,\rm{nm}$ subjected to annealing of different duration as indicated in the legend. (b) is a graph of the measured normalized resistance as a function of temperature for different annealing duration. The inset shows the SEM image of a device, the white bar corresponds to 1.5$\,\mu$m. (c-d) are summary plots of the critical temperature and critical current for the Hall bars with $\rm{W}\,=\,50\,\rm{nm}$ as a function of annealing time, respectively. (e) Shows a plot of the temperature dependence of the normalized resistance of a large MoRe structure reported in the inset after different annealing times.}
\end{figure}

In order to test the suitability of MoRe for nanometer scale devices, we fabricated structures with channel width of $50\,\rm{nm}$ and characterize their superconducting properties as a function of annealing time. These narrow channel devices are also superconducting and have large values of I$_{\rm{C}}\,\approx\,50\,\rm{\mu A}$ at $\rm{T}\,=\,4.2\,\rm{K}$, see Figure 2a. After annealing the devices at $400\,^{\circ}\mathrm{C}$ the critical current increases, reaching values as high as $0.2\,\rm{mA}$, following the trend previously reported for $\rm{W}\,=\,200\,\rm{nm}$. Upon annealing, the superconducting critical temperature  increases and the normalized resistance \textit{versus} temperature shows a sharp transition, indicating a uniformly defined superconducting state (see Figure 2b). A plot of T$_{\rm{C}}$ and I$_{\rm{C}}$ as a function of the annealing time shows that both superconducting parameters increase upon increasing the annealing time, see Figure 2c-d. 

The superconducting properties of MoRe alloys are known to depend on various factors, such as (1) composition of the material \cite{F. J. Mori1963}, (2) film thickness (e.g. $6\,\rm{\mu m}$ thick Mo$_{50}$Re$_{50}$ film has T$_{\rm{C}}\,=\,15\,\rm{K}$ \cite{J. R. Gavaler1972}, whereas T$_{\rm{C}}$ of Mo$_{60}$Re$_{40}$  increases from 5.6 to $9.7\,\rm{K}$ increasing the film thickness from $2\,\rm{n}$m to $10\,\rm{nm}$ \cite{V. A. Seleznev2008}) and (3) film homogeneity (annealing at temperatures higher than $750\,^{\circ}\mathrm{C}$ for $15\,\rm{min.}$ increases T$_{\rm{C}}$ by $\sim\,0.5\,\rm{K}$ \cite{S.M. Deambrosis2006}). Since in our experiments we only vary the annealing time, the observed improvement of the superconducting properties of the studied alloy of MoRe could in principle be the consequence of changes in the MoRe crystal structure or desorption of contaminants from the metal. 

To discern the leading mechanism (i.e. desorption or recrystallization) responsible for the improved superconductivity with low-temperature annealing, we prepared large devices of $250\,\rm{\mu m}$ width (shown in the inset of Figure 2e). In this case we find that the superconducting transition occurs at T$_{\rm{C}}\,=\,7.5\,\rm{K}$ independently of the annealing time and this is characterized by a sharp drop of resistance indicating a uniformly defined superconducting state. The fact that the superconducting properties of large samples do not change upon annealing excludes the possibility that the chosen annealing procedure changes the crystal structure of MoRe. Indeed, if annealing was responsible of a recrystallization of MoRe, we would have observed a similar dependence of the superconducting properties upon annealing independently of the superconducting channel width, and this is clearly not the case in our experiments. On the other hand, the fact that in wide superconducting channels the superconducting properties do not change with annealing indicates that desorption of contaminants from the metal edges is the dominant factor. Indeed, annealing at $400\,^{\circ}\mathrm{C}$ in constant flow of Ar/H$_{2}$ gasses is known to remove PMMA residuals efficiently also from graphene.  

\begin{figure}[ht]{}
\includegraphics[width=0.45\textwidth]{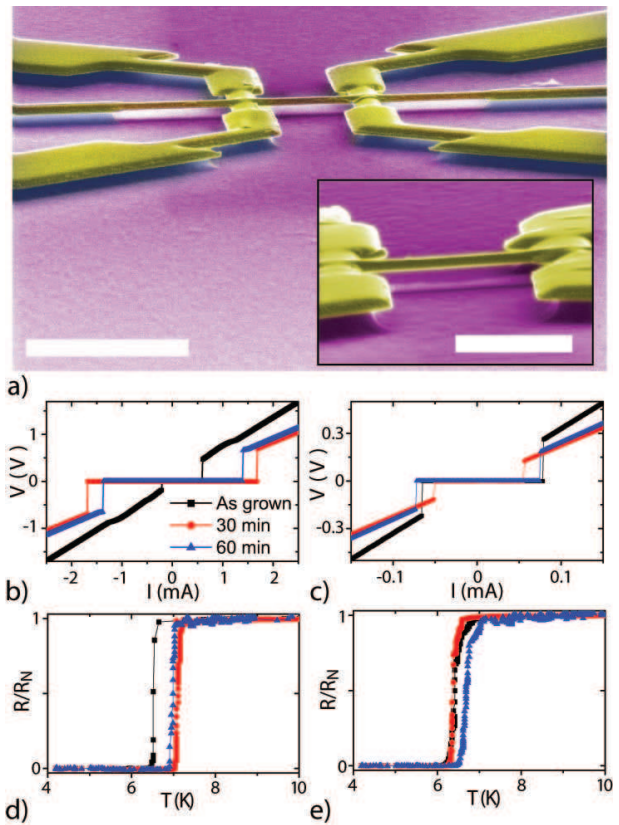}
\caption{\label{fig3} Panel (a) shows false coloured SEM images of suspended MoRe with $\rm{W}\,=\,50\,\rm{nm}$. The white scale bar corresponds to $3\,\rm{\mu m}$ for the main figure and $500\,\rm{nm}$ for the image in the inset. (b) and (c) are graphs of the measured V-I characteristics for suspended MoRe devices with $200\,\rm{nm}$ and $50\,\rm{nm}$ width respectively and for different annealing duration as indicated in the legend of (b). (d) and (e) show the measured normalized temperature dependence of the resistance in suspended MoRe devices with $200\,\rm{nm}$ and $50\,\rm{nm}$ width respectively and for different annealing duration as indicated in the legend of (b).}
\end{figure}

Finally, contrary to other commonly used superconductors for nanoscale devices (e.g. Al, Nb, etc..), MoRe is not etched by Hydro-fluoric acid (HF). Hence, we set to demonstrate the suitability of this alloy for the fabrication of suspended superconducting nanochannels ($\rm{W}\,=\,200\,\rm{nm}$ and $50\,\rm{nm}$) using standard HF wet-etching of SiO$_2$ sacrificial layer. More than 50 suspended superconducting bridges have been fabricated using the aforementioned steps, see Figure 3a. Representative V-I curves and temperature dependence of the normalized resistance measured in as-grown and annealed suspended devices are shown in Figure 3b-e. In all cases we find that I$_{\rm{C}}$ and T$_{\rm{C}}$ improve upon annealing, similarly to the observations for the supported devices. More specifically, T$_{\rm{C}}$ measured in devices with $\rm{W}\,=\,200\,\rm{nm}$ increases from $6.4\,\rm{K}$ in as-grown to $7\,\rm{K}$ after $60\,\rm{min}$ annealing, see Figure 3d. Narrower structures with $\rm{W}\,=\,50\,\rm{nm}$ also show an increase of T$_{\rm{C}}$ from $6\,\rm{K}$ to $6.4\,\rm{K}$ upon annealing, see Figure 3e. The experimental finding that MoRe is highly compatible to the standard annealing procedure performed on graphene, makes this material potentially interesting for exploring superconducting electromechanical structures with carbon-based nanomaterials such as carbon nanotubes \cite{K. A. Huttel2009} and graphene.  

In summary we have demonstrated the successful fabrication of superconducting nanostructures of MoRe alloy using standard PMMA and lift-off procedure, with critical temperature as high as $6.5\,\rm{K}$ in channels of $50\,\rm{nm}$ width.  An annealing procedure commonly used in the fabrication of high quality graphene transistors, is employed to desorb contaminants of PMMA from the MoRe and improve the superconducting properties such as critical temperature and critical current. Finally we demonstrate the fabrication of suspended MoRe nanostructures with channel width of $200\,\rm{nm}$ and $50\,\rm{nm}$ and channel length up to $3\,\rm{\mu m}$ using standard HF wet-etching of the SiO$_2$ substrate. These suspended structures show similar values of critical temperature and critical current to those found in supported devices. These findings pave the way for the developement of nanoscale superconducting electromechanical devices using standard fabrication techniques.

SR acknowledges financial support from EPSRC (Grant no. EP/J000396/1, EP/K017160/1 and EP/K010050/1).

\end{document}